\documentclass{PoS}

\usepackage{amsmath}
\usepackage{graphicx}
\usepackage{xcolor}

\title{Chiral restoration and deconfinement in two-color QCD with two flavors of staggered quarks}

\ShortTitle{Chiral restoration and deconfinement in QC$_2$D with two flavors of staggered quarks}

\newcommand{\affDARMSTADT}{Institut f\"ur Kernphysik, Technische Universit\"at Darmstadt, 64289 Darmstadt, Germany}
\newcommand{\affGIESSEN}{Institut f\"ur Theoretische Physik, Justus-Liebig-Universit\"at, 35392 Gie{\ss}en, Germany}
\newcommand{\affBIELEFELD}{Fakult\"at f\"ur Physik, Universit\"at Bielefeld, 33501 Bielefeld, Germany}

\author{\speaker{David Scheffler}\\
        \affDARMSTADT\\
        E-mail: \email{dscheff@theorie.ikp.physik.tu-darmstadt.de}}

\author{Christian Schmidt\\
       \affBIELEFELD\\
       E-mail: \email{schmidt@physik.uni-bielefeld.de}}

\author{Dominik Smith\\
       \affDARMSTADT\\
       E-mail: \email{smith@theorie.ikp.physik.tu-darmstadt.de}}

\author{Lorenz von Smekal\\
       \affDARMSTADT\\
       \affGIESSEN\\
       E-mail: \email{lorenz.smekal@physik.tu-darmstadt.de}}

\abstract{In preparation of lattice studies of the two-color QCD phase
  diagram we study chiral restoration and deconfinement at finite
  temperature with two flavors of staggered quarks using an RHMC
  algorithm on GPUs. We first study unquenching effects in local Polyakov
  loop distributions, and the Polyakov loop potential obtained via
  Legendre transformation, in a fixed-scale approach for heavier
  quarks. We also present the chiral condensate and the corresponding
  susceptibility over the lattice coupling across the chiral
  transition for lighter quarks. Using Ferrenberg-Swendsen reweighting
  we extract the maxima of the chiral susceptibility in order to determine
  pseudo-critical couplings on various lattices suitable for chiral
  extrapolations. These are then used to fix the relation between coupling and
  temperature in the chiral limit.}  

\FullConference{31st International Symposium on Lattice Field Theory LATTICE 2013\\
                July 29 - August 3, 2013\\
                Mainz, Germany}

\begin{document}

\section{Introduction}
  Direct calculations using standard Monte-Carlo methods in quantum chromodynamics (QCD) at finite baryon density are hindered by the fermion sign problem. In order to still gain insight into gauge theories at finite densities several theories free of the sign problem have been investigated like two-color QCD \cite{Hands:2000ei,Cotter:2012mb}, adjoint QCD \cite{Karsch:1998qj,Hands:2000ei} or $G_2$-QCD \cite{Maas:2012wr,vonSmekal:2013qqa}.
This work deals with two particular aspects in two-color QCD at finite temperature: the determination of effective Polyakov loop potentials and the study of chiral properties unique to staggered fermions. Here we only present calculations at $\mu=0$ leaving the inclusion of finite density for upcoming work.

The interest in effective Polyakov loop potentials mainly arises from Polyakov-loop enhanced effective models, like PQM or PNJL models, which use Polyakov loop potentials as an essential input. Little is known about its density dependence, but a good description of the two-color QCD phase diagram seems possible if it were \cite{Strodthoff:2013cua}.    
In contrast to other works where the potential of the volume-averaged Polyakov loop has been studied \cite{Fischer:2013eca} we here use the local (site-wise) effective potential.
In pure gauge this has recently been investigated~\cite{DSmith-2013}. We will build on this work by including rather heavy quarks in the staggered formulation in order to assess unquenching effects.

Two-color QCD at vanishing density has a flavor symmetry which is enlarged from the usual $SU(N_f)_L \times SU(N_f)_R \times U(1)_B$ to $SU(2 N_f)$ (Pauli-G\"ursey symmetry). A chiral condensate or a Dirac mass term break this down to $Sp(N_f)$ in the continuum. In contrast, due to the missing Dirac structure of the staggered discretization, one expects a breaking to $SO(2 N_f)$ in the continuum limit for staggered quarks. In our case of $N_f=2$ the expected breaking pattern in the continuum limit thus is $SU(4) \simeq  SO(6) \rightarrow SO(4)$ (whereas for fundamental Dirac fermions in the continuum one would have: $SU(4) \simeq SO(6) \rightarrow Sp(2)\simeq SO(5)$). 
We are interested in the scaling behavior of this atypical breaking and determine the lattice scale via (pseudo-)critical couplings.

We choose the Wilson gauge action for $SU(2)$ and utilize the standard staggered fermion discretization with the rooting procedure to describe two flavors of quarks in the fundamental representation. Configurations are generated using RHMC.

\section{Effective Polyakov loop potentials}
  Our calculations for evaluating the Polyakov loop effective potential are performed in the fixed scale approach. At fixed values of the coupling $\beta$ we vary the temporal lattice extent $N_t$ in order to change the temperature according to
\begin{equation}
	T(N_t)=\frac{1}{a N_t} \text.
\end{equation}
Critical couplings, the lattice spacing and the temperature scale for $SU(2)$ have been precisely determined in pure gauge simulations~\cite{DSmith-2013}.
We employ the same values of the coupling, $\beta_{10}=2.577856$ and $\beta_{12}=2.635365$, which are the critical couplings in pure gauge for $N_t=10 \text{ and } 12$ lattices. The scale $a \sqrt{\sigma}$ amounts to $0.140$ at $\beta_{10}$ and $0.116$ at $\beta_{12}$.
In our study we use lattices with temporal extent $N_t=6,8,10 \text{ and }12$ with an aspect ratio $N_s/N_t$ of $4$. Finite volume tests with an aspect ratio of $3$ have been performed for some lattices and no difference in observables (average Polyakov loop and chiral condensate) as well as in the local Polyakov loop distributions have been visible.
Relative temperatures have been obtained via
\begin{equation}
\left.\frac{T(N_t)}{T_c}\right|_\beta=\frac{N^c_t}{N_t}	\text.
\end{equation}
The requirement of fixed physical quark masses leads to rescaled quark masses in lattice units at the different values of $\beta$:
\begin{equation}
 am(\beta')= \frac{a(\beta')}{a(\beta)} \cdot am(\beta)
 \label{eq:masses}
\end{equation}
Masses have been chosen to be $a m = 0.1 \text{ and } 0.5$ at $\beta_{10}$ which via Eq.~\eqref{eq:masses} results in $a m = 0.083 \text{ and } 0.414$ for $\beta_{12}$.
Introducing fermions will modify the lattice scale and it would be necessary to determine the scale at the different quark masses. As our quark masses are rather large, we assume that the scale does not change much. At least qualitatively the results should not change.

For the calculation of the local effective Polyakov loop potential we need the Polyakov loop probability distribution $P(l)$. It is obtained via histogramming the Polyakov loop values on the spatial lattice volume and aggregating over the whole ensemble.

From the distribution we first calculate the ``constraint'' effective potential $V_0$
\begin{equation}
		V_0(l) = -\log P(l) \text{.}
\end{equation}
Next we perform a Legendre transform as follows:
\begin{equation}
			W(h) = \log \int dl \exp{\left(-V_0(l) +h l\right)}\quad\quad
			V_{\textrm{eff}}(\hat l) = \sup_h{\left(\hat l h - W(h)\right)}\text.
\end{equation}
Its minimum is correctly located at the value of the average Polyakov loop.

At the critical temperature $T_c$ the constraint Polyakov loop potential in pure gauge is given by the Vandermonde potential and corresponding distribution
\begin{equation}
	V_0^{(T_c)}(l)=-\frac{1}{2} \log(1-l^2) - C\quad\quad\quad\quad\quad\quad
	P^{(T_c)}(l) = \frac{2}{\pi} \sqrt{1-l^2} \text.
	\label{eq:vandermonde}
\end{equation}
An ansatz for temperatures above $T_c$ is given by \cite{Smith:2009kp}
\begin{equation}
	V_0(l) = V_0^{(T_c)}(l) + a(T) - b(T) l + c(T) l^2
	\label{eq:ansatz_pot}
\end{equation}
corresponding to
\begin{equation}
	P(l) = \frac{2}{\pi} \sqrt{1-l^2} \exp{\left(-a(T) +b(T) l -c(T) l^2\right)} \text.
	\label{eq:ansatz_dist}
\end{equation}

\begin{figure}
	\includegraphics[width=.5\linewidth]{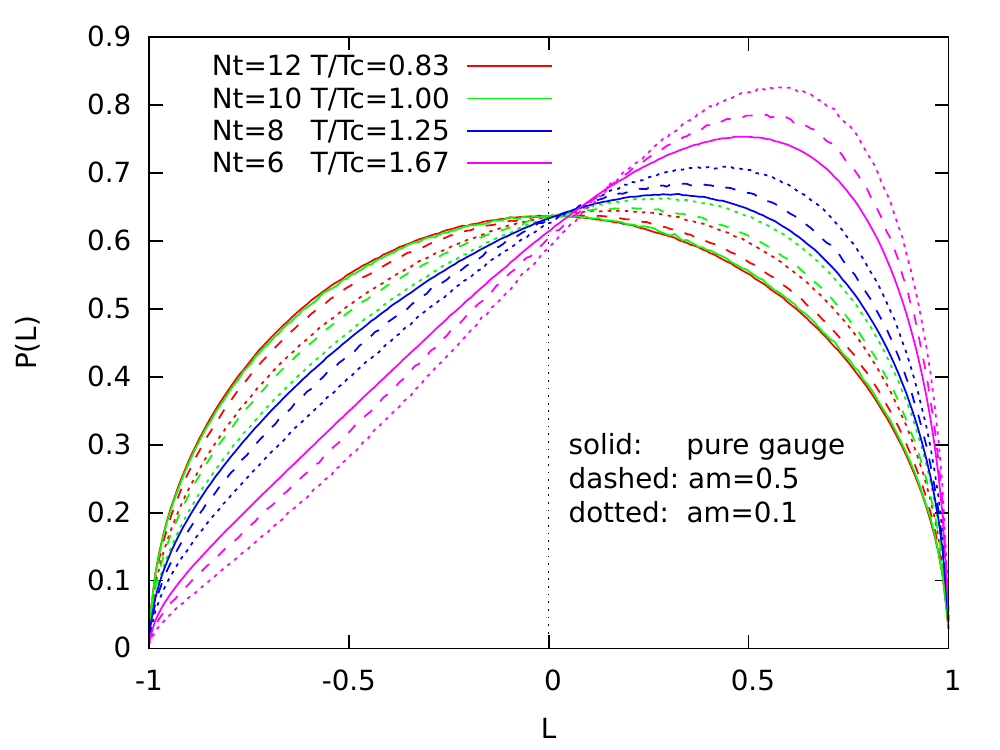}
	\includegraphics[width=.5\linewidth]{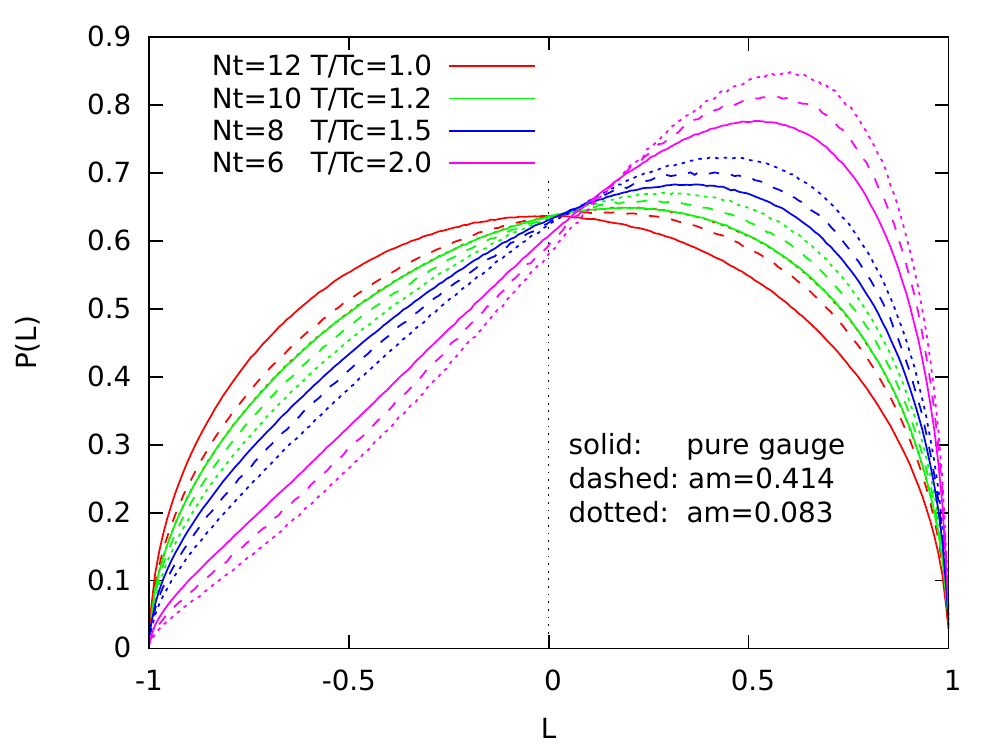}
	\caption{\label{dist}Distribution of the SU(2) Polyakov loop at $\beta=2.577856$ (left) and $\beta=2.635365$ (right)}
\end{figure}

Figure~\ref{dist} shows the Polyakov loop distributions for different temperatures comparing pure-gauge results to unquenched ones with two different quark masses.
For temperatures at and below the critical temperature in pure gauge the distribution and correspondingly the constraint effective potential (not shown) are perfectly described by Eq.~\eqref{eq:vandermonde}. $Z(2)$ symmetry is spontaneously broken at temperatures above $T_c$, leading to an asymmetry towards positive Polyakov loop values in the distribution and the constraint effective potential. The unquenched simulations show a stronger asymmetry in these observables than pure gauge caused by the explicit breaking of the $Z(2)$ symmetry by the finite fermion masses. Explicit symmetry breaking and thus the asymmetry is larger for smaller quark masses.
In the effective potentials (not shown) the breaking manifests itself in a shift of the potential minimum towards larger Polyakov loop values.

\begin{figure*}
	\includegraphics[width=0.33\linewidth]{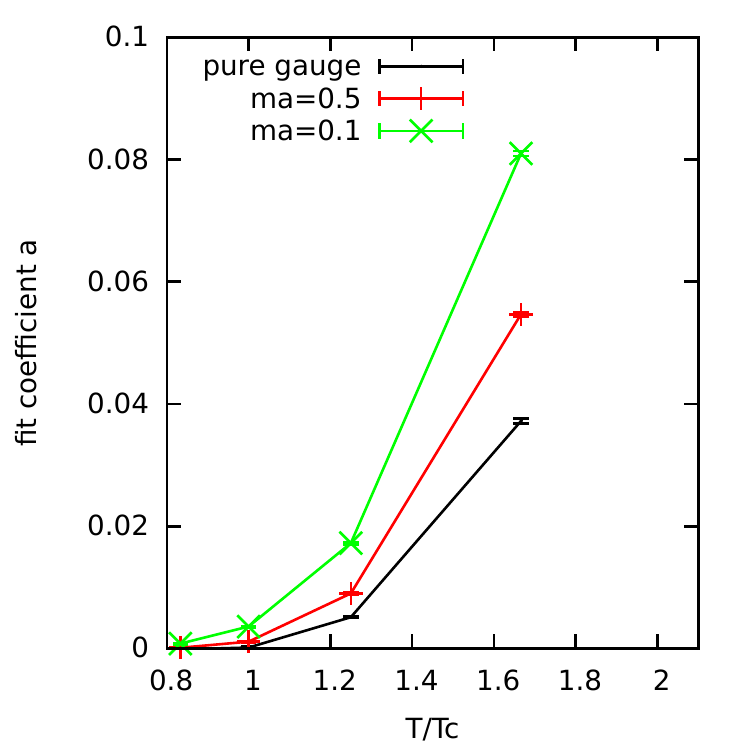}
	\includegraphics[width=0.33\linewidth]{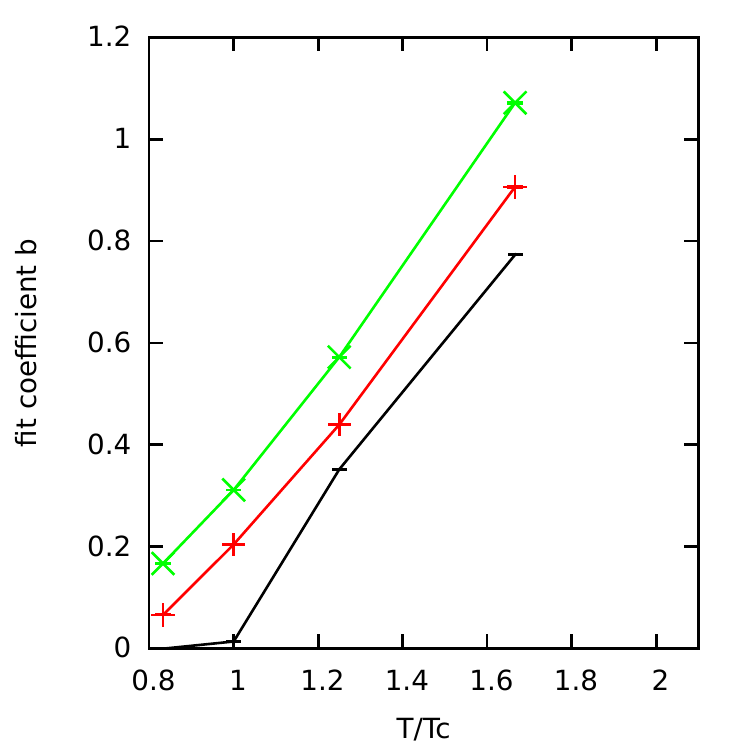}
	\includegraphics[width=0.33\linewidth]{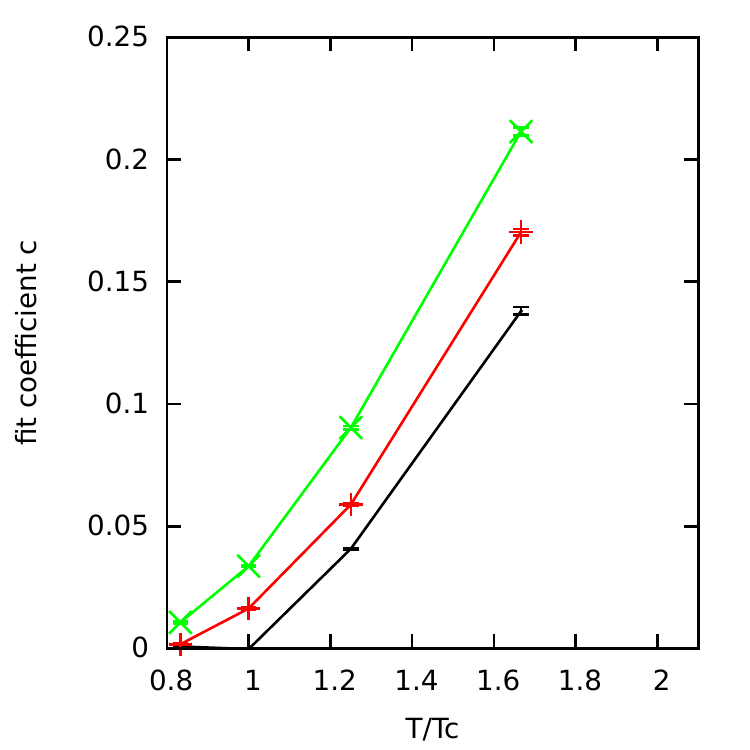}
	\caption{\label{coeff2.5}Temperature dependence of fit coefficients for the Polyakov loop distribution at $\beta=2.577856$}
\end{figure*}

The temperature-dependent coefficients in Equations \eqref{eq:ansatz_pot} and \eqref{eq:ansatz_dist} are obtained from a $\chi^2$ fit. 
Fig.~\ref{coeff2.5} shows results for $\beta=2.577856$. We find that all three parameters rise with increasing temperature and also with decreasing mass. Note that $b(T)$ amounts to the breaking of $Z(2)$ center symmetry. This coefficient shows a qualitative difference between the pure gauge results and those with dynamical quarks included: While the coefficient is consistent with zero at and below $T_c$ in the pure gauge case, it no-longer drops to such small values near $T_c$ when dynamical quarks are included due to their explicit symmetry breaking.
More details will be reported in \cite{DS_effpol_prep}.

\section{Chiral properties}
  For the analysis of chiral properties we have not used the fixed-scale approach but varied $\beta$ to change the temperature. The fixed-scale approach is limited by the number of temperature values inside the interesting temperature range. 
We have performed simulations on various lattice sizes $N_t=4, 6, 8$ with an aspect ratio $N_s/N_t=4$, and various fermion masses $am$ between $0.005 \text{ and } 0.1$.

\begin{figure}
	\centering
	\includegraphics[width=0.48\textwidth]{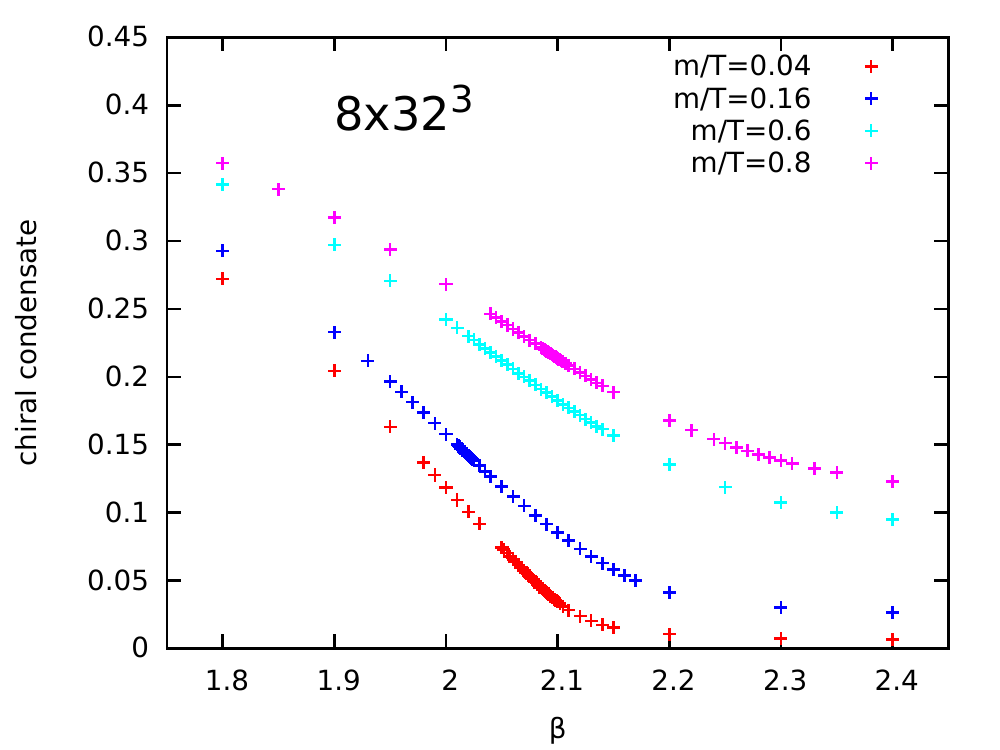}
	\includegraphics[width=0.48\textwidth]{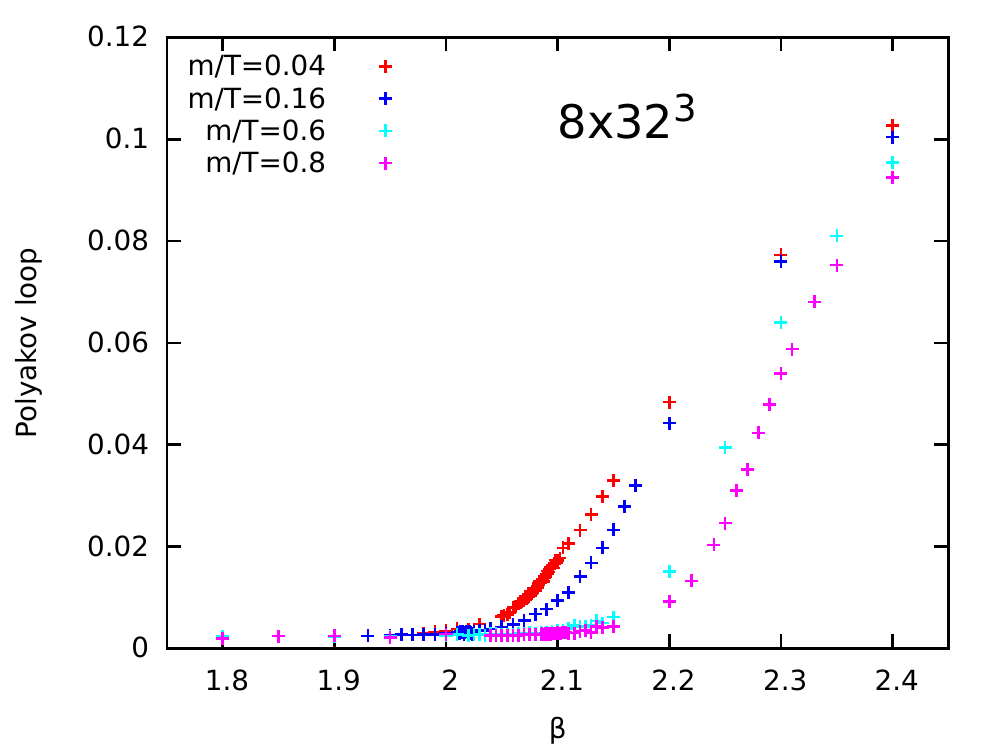}
	\caption{\label{orderparams8}Chiral condensate (left) and Polyakov loop (right) as function of the coupling for different masses}
\end{figure}

We first present simple observables, namely the chiral condensate and the Polyakov loop. Figure~\ref{orderparams8} shows their dependence on the coupling and the quark mass on $N_t=8$ lattices. Errors are of the size of the symbols or smaller and are omitted for better visibility. We clearly find a crossover transition in both pseudo-order parameters for all masses. For lower quark masses the transition is stronger (steeper) in the chiral condensate and weaker in the Polyakov loop and vice versa which is the expected behavior. The chiral condensate at large couplings is still large for $m/T=0.6\text{ and }0.8$ due to large explicit symmetry breaking, but drops to low values for $m/T=0.04\text{ and }0.16$.

\begin{figure}
	\includegraphics[width=0.33\textwidth]{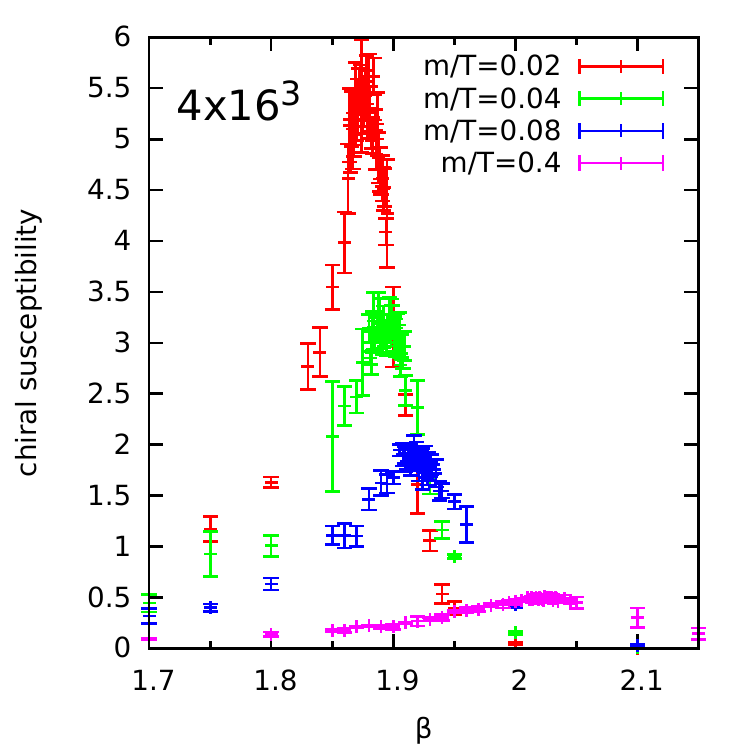}
	\includegraphics[width=0.33\textwidth]{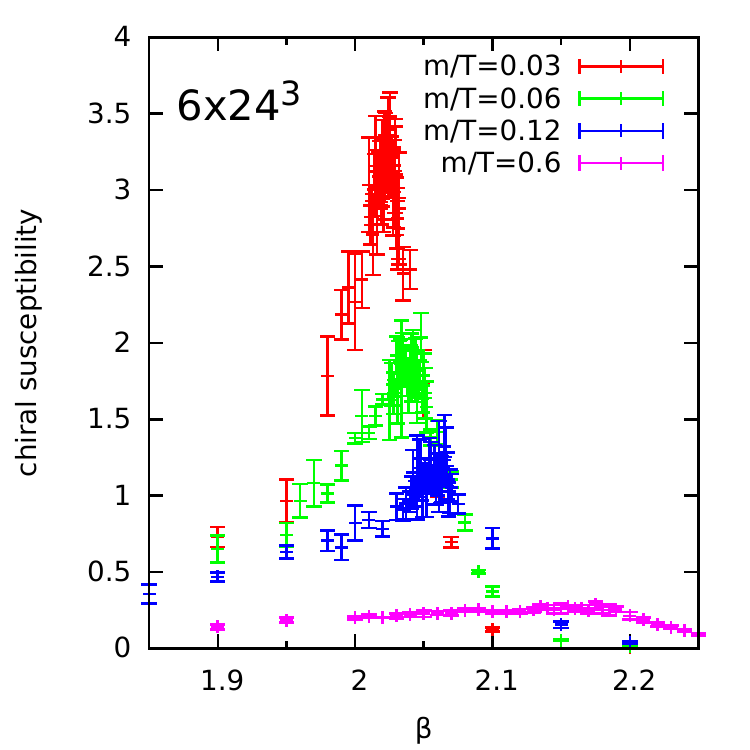}
	\includegraphics[width=0.33\textwidth]{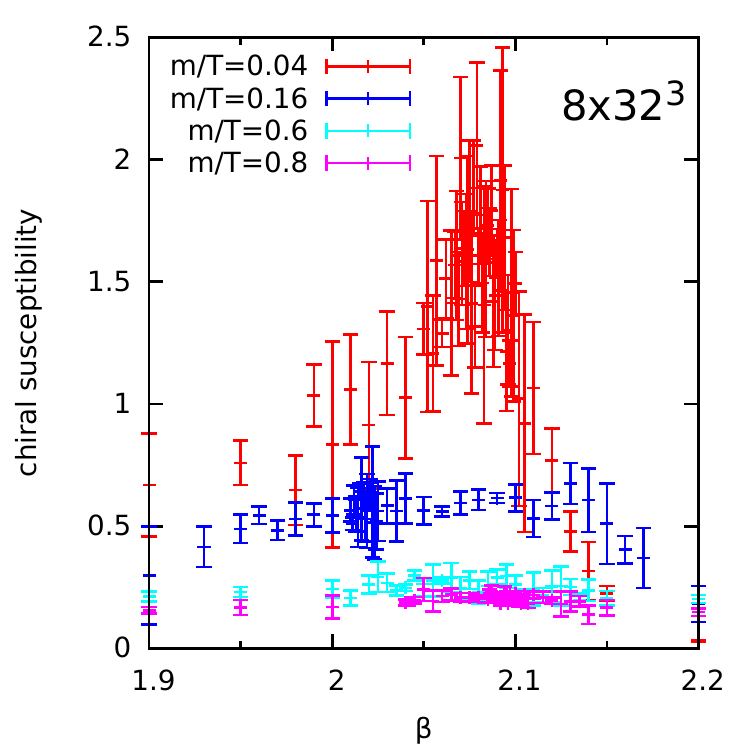}
	\caption{\label{susc}Chiral susceptibility (disconnected part) at different masses and lattice sizes}
\end{figure}

\begin{figure}
	\includegraphics[width=0.5\textwidth]{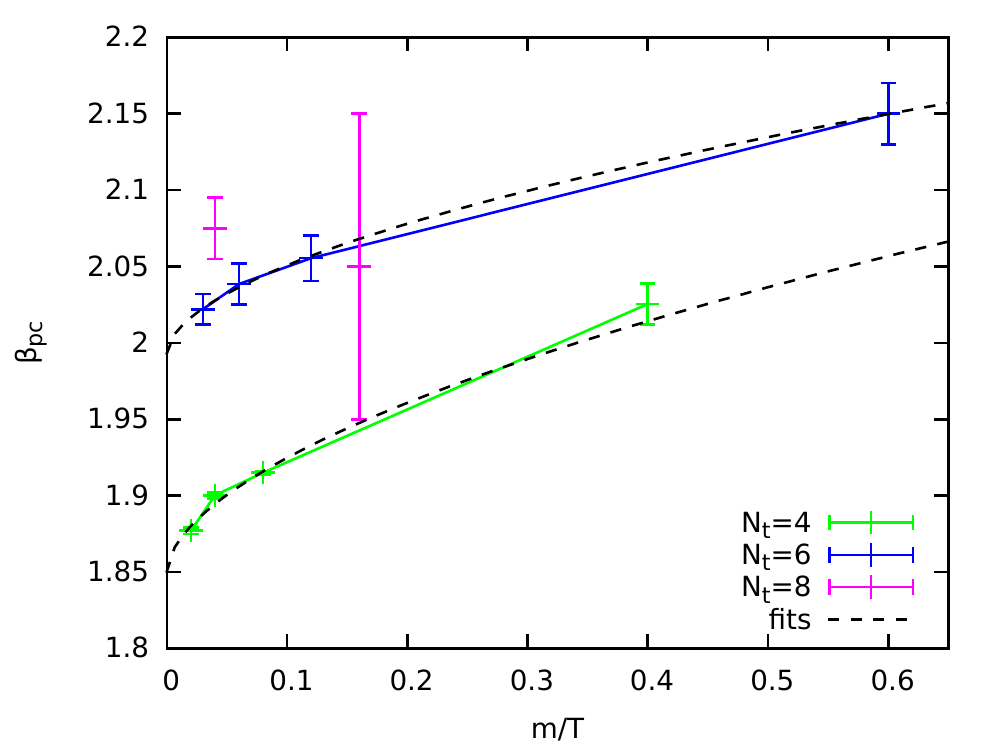}
	\includegraphics[width=0.5\textwidth]{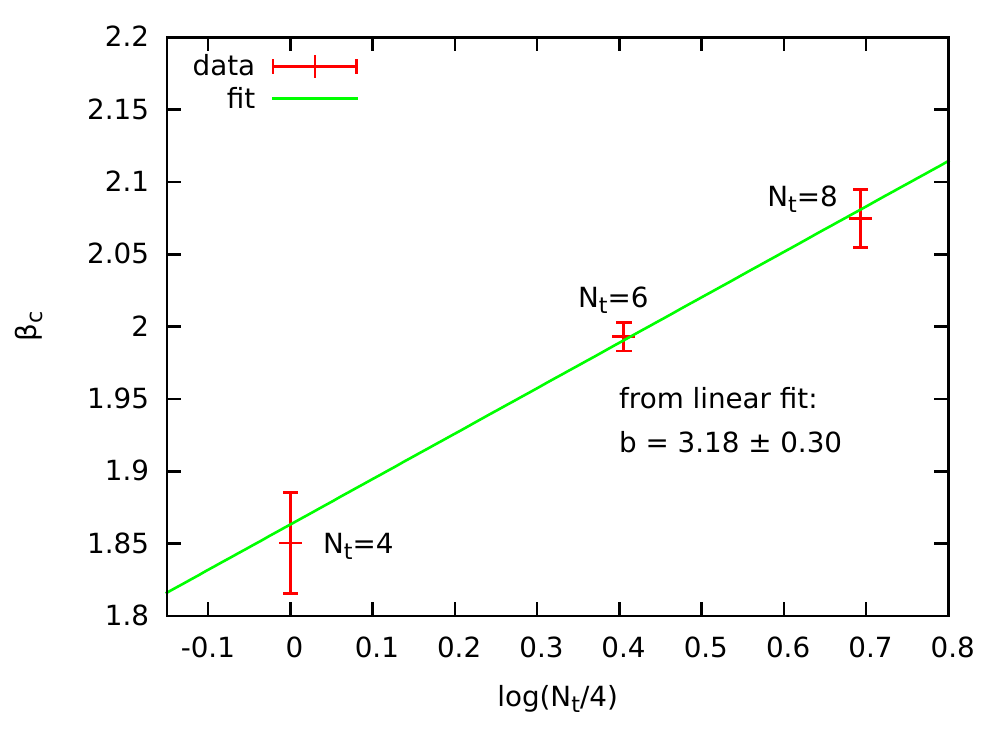}
	\caption{\label{tempscale}Pseudo-critical couplings (left) and fit for temperature scale (right)}
\end{figure}

The disconnected part of the chiral susceptibility for several quark masses and lattice sizes is given in Figure~\ref{susc}. For small quark masses clear peaks are visible. Larger sample sizes are needed for most calculations at $N_t=8$ to reduce statistical errors.
From the location of the peaks of the chiral susceptibility we obtain the pseudo-critical couplings as function of temporal lattice extent and quark mass. 
On lattices with temporal extent of $4$ and $6$ the peak position can readily be determined. For the $N_t=4$ results we additionally utilize Ferrenberg-Swendsen reweighting to obtain an even more precise location of the maxima.
Unfortunately, on the $N_t=8$ lattices the peak position is not as clear due to larger statistical uncertainties. In the case of $m/T=0.8 \text{ and } 0.6$ we find very broad and flat maxima and thus cannot determine a peak position. Also at $m/T=0.16$ the uncertainty is so large that we will not include this data point in the determination of the scale.
Resulting pseudo-critical couplings are shown in Figure~\ref{tempscale} (left).
We extrapolate the pseudo-critical couplings to the chiral limit, $m=0$, separately for $N_t=4 \text{ and }6$ using the ansatz
\begin{equation}
	\beta_{pc}(m,N_t)=\beta_c(N_t)+d\cdot (m/T)^c \text{.} \label{betac}
\end{equation}
For $N_t=8$ we use the pseudo-critical coupling of the smallest mass value as our currently best estimate for the critical coupling.

%
Using these leading scaling behavior fits, we can relate the extrapolated critical couplings for different temporal lattice sizes to a relative temperature scale. At fixed $N_t$ we first assume that we may write, in the chiral limit,  
\begin{equation}
	\frac{T}{T_c}=\frac{a_c}{a}=\exp\big\{b(\beta-\beta_c)\big\} \text{,} 
\label{eq:scale_fixednt} 
\end{equation}
for temperatures $T$ sufficiently close to $T_c$, where $b$ is a
nonperturbative coefficient that we wish to estimate. To do that we
compare criticality for different $N_t$,  
\begin{equation}
	1 = \frac{N_t^{(1)} a_c^{(1)}}{N_t^{(2)} a_c^{(2)}} \quad\Leftrightarrow\quad \frac{a_c^{(1)}}{a_c^{(2)}}  = \frac{N_t^{(2)}}{N_t^{(1)}}  = \exp\big\{b(\beta_c^{(2)} -\beta_c^{(1)} )\big\}  \text{,} \label{eq:scale_temp}  
\end{equation}
with the same coefficient $b$ (the lattice spacing $a_c^{(2)} $ and coupling $\beta_c^{(2)} $ are not critical on the $N_t^{(1)}$  lattice for which we need their leading logarithmic dependence in \eqref{eq:scale_fixednt}).
From the three different critical couplings at $N_t=4, 6$ and $8$ 
we obtain $b=3.18 \pm 0.30$, compare Figure~\ref{tempscale}
(right). For comparison, from the 
deconfinement phase transition in the pure gauge theory the
corresponding value has incidently been determined to be $b=3.28 \pm
0.07$ \cite{DSmith-2013}.

Close to criticality we furthermore expect magnetic scaling. The
exponent $c$ in our fit ansatz (\ref{betac}) for example, with
(\ref{eq:scale_fixednt}) at leading order in the critical temperature
$t = T/T_c -1 $ determines the variation of the location
$t_\mathrm{max}(m) \sim m^{1/(\beta\delta)} $ of the 
susceptibility maxima with the explicitly symmetry breaking
$m$ \cite{Pelissetto:2000ek}, hence
$c^{-1} = \beta\delta $. A global fit to our  
$N_t=4 \text{ and }6$ data yields $c=0.6(2)$ as our currently best
estimate for the expected $SO(6) \to SO(4)$ universality class.  

In order to determine the critical exponents $\beta$ and $\delta$, and
with two exponent-scaling the remaining ones from scaling and
hyperscaling relations as well, we need a second estimate, however.

One candidate is the height $\chi_\mathrm{max}(m) $ of the peak in the 
chiral susceptibility, whose leading behavior as function of the quark
mass around criticality is given by
\begin{equation}
	\chi_\mathrm{max}(m)/T^2 = \left.\chi(m)\right|_{\beta_{pc}(N_t,m)}/T^2 \sim (m/T_c)^{1/\delta-1}\text{.} \label{delta_chimax}
\end{equation}
Another possibility is to investigate the mass dependence of the
chiral condensate  right at the critical coupling, 
\begin{equation}
	m \left.\left< \bar\psi \psi\right>(m)\right|_{\beta_c(N_t)}/T_c^4 \sim (m/T_c)^{1/\delta+1}\text{,} \label{delta_chcond}
\end{equation}
Both are governed by the same critical exponent
$\delta$. Unfortunately, the accuracy of our present determinations of
the susceptibility maxima is not yet sufficient to obtain consistent
results. From our currently best analysis with Ferrenberg-Swendsen reweighting 
of the  $N_t=4$ lattice data, at the moment we obtain $\delta=3.4(3)$
with Eq.~\eqref{delta_chimax} as compared to  $\delta=4.25(3)$
from Eq.~\eqref{delta_chcond}.

\section{Summary and outlook}
  We have assessed the unquenching effects of quarks on the effective Polyakov loop potential. Regarding the chiral sector we have started the determination of a relative scale. Further simulations at $\mu=0$ are ongoing to improve this determination. They will hopefully allow us to also obtain more reliable results for the  critical exponents eventually. After analyzing two-color QCD along the temperature axis at $\mu=0$ we will proceed to finite chemical potential and investigate the influence of finite density on the effective Polyakov loop potential for a start. 

\section*{Acknowledgements}
This work was supported by the BMBF under contract 06DA9047I, by the Deutsche Forschungsgemeinschaft within SFB 634, by the Helmholtz International Center for FAIR within the LOEWE initiative of the State of Hesse and by the European Commission, FP7-PEOPLE-2009-RG, No. 249203.

\providecommand{\href}[2]{#2}\begingroup\raggedright\endgroup

\end{document}